\documentclass[aps,reprint,pre,showpacs,superscriptaddress]{revtex4-2}
\usepackage{amsmath}
\usepackage{amssymb}
\usepackage{graphicx}
\usepackage{color}
\usepackage{bm}
\usepackage{siunitx}
\usepackage{mathtools}
\usepackage{MnSymbol}
\usepackage{cancel}
\usepackage{dcolumn} % align on decimal point in tables

\usepackage{babel}

\definecolor{linkcolor}{rgb}{0.56640625,0.3359375,0.42578125} % définition de la couleur des liens pdf

\usepackage[pdftex,colorlinks=true,
pdfstartview=FitV,
linkcolor= linkcolor,
citecolor= linkcolor,
urlcolor= linkcolor,
hyperindex=true,
hyperfigures=false]
{hyperref} % fichiers pdf 'intelligents', avec des liens entre les références, etc.

\begin{document}

\title{Protocol dependence for avalanches under constant stress in elastoplastic models}

\author{Tristan Jocteur}
\author{Eric Bertin}
\author{Romain Mari}
\author{Kirsten Martens}
\affiliation{Univ. Grenoble-Alpes, CNRS, LIPhy, 38000 Grenoble, France}

\begin{abstract}
Close to the yielding transition, amorphous solids exhibit a jerky dynamics characterized by plastic avalanches. The statistics of these avalanches have been measured experimentally and numerically using a variety of different triggering protocols, assuming that all of them were equivalent for this purpose. In particular two main classes of protocols have been studied, deformation under controlled strain or under controlled stress. In this work, we investigate different protocols to generate plasticity avalanches and conduct twodimensional simulations of an elastoplastic model to examine the protocol dependence of avalanche statistics in yield-stress fluids. We demonstrate that when stress is controlled, the value and even the existence of the exponent governing the probability distribution function of avalanche sizes strongly depend on the protocol chosen to initiate avalanches.
This confirms in finite dimensions a scenario presented in a previous mean-field analysis. 
We identify a consistent stress-controlled protocol whose associated avalanches differ from the quasi-static ones in their fractal dimension and dynamical exponent. Remarkably, this protocol also seems to verify the scaling relations among exponents previously proposed. Our results underscores the necessity for a cautious interpretation of avalanche universality within elastoplastic models, and more generally within systems where several control parameters exist.
\end{abstract}

\maketitle

\section{Introduction}

In both nature and technical applications, one frequently encounters the phenomenon of intermittent dynamics characterized by scale-free event statistics. Avalanches involve strongly correlated dynamics of complex geometrical objects across scales ranging from atomic to tectonic. Examples include avalanches in magnetic materials \cite{Laurson2014}, superconductors \cite{Verma2016}, glasses \cite{Antonaglia2014}, irreversible rearrangements in soft matter systems \cite{cantat2005dissipative}, critical dynamics of imbibition fronts \cite{Clotet2016} and crack growth \cite{Kokkoniemi2017}, and the mechanical responses of granular and porous media \cite{Main2017}, wood \cite{Makinen2015}, and geological flows like snow avalanches \cite{VanHerwijnen2016}, landslides, and earthquakes \cite{DeArcangelis2016}.

Operating near criticality can be beneficial in certain natural systems, as seen in avalanche dynamics within neural networks \cite{dearcangelisActivityDependentNeuronalModel2012} and the collective movement of fish schools \cite{puy2023self}. However, intermittent dynamics can also pose significant challenges, such as catastrophic events in form of earthquakes or undesired phenomena in engineering applications, such as brittle failure due to plastic avalanches or stick-slip motion of frictional surfaces.

Understanding the complex, nonlinear spatio-temporal responses of these systems and their connections across different scales is essential for accurate physical predictions and the development of reliable geo-physical and engineering models. In some cases the scale free dynamics can be associated with out-of-equilibrium phase transitions, such as the depinning transition \cite{kolton2006dynamics} and the yielding transition \cite{lin_scaling_2014}. Key questions in this field revolve around understanding the emergence of universality \cite{Denisov2016} and the potential for defining various universality classes \cite{lin_scaling_2014, Liu2016,jocteurYieldingAbsorbingPhase2024a}, which has initiated considerable debate in recent years \cite{Ispanovity2014}.

While the detailed study of non-universal features in avalanche dynamics has not been extensively covered, our work aims to address this gap. Previous studies have explored the effects of inertia \cite{Salerno2013, Karimi2017} and transient dynamics \cite{Leishangthem2017}, but research on the impact of different protocols for triggering avalanches remains limited. 

The theoretical work by Pérez-Reche et al.~\cite{perez2008driving} on a random soft spin model demonstrates the possible difference between avalanches occurring in strain- versus stress-controlled protocols. Such a difference has been experienced also in experiments on martensitic materials by Vives et al.~\cite{vives2009driving}.
% Eduardo 
Jagla further pointed out that, within a mean-field model for elastoplastic dynamics associated with the yielding transition, one expects an additional dependence on the way avalanches are triggered~\cite{jagla_avalanche-size_2015}.

In this manuscript we investigate the protocol dependence of avalanche dynamics within a spatially resolved elastoplastic description of the yielding transition. 
This allows not only to test existing mean-field predictions, but also to establish a more comprehensive understanding of the spatial complexity made apparent in the fractal dimension of the avalanches occurring under various protocols.

Our manuscript is organised as follows. In Sec.~\ref{sec:model:protocols}
we introduce the twodimensional elastoplastic model used for simulations with strain or stress controlled dynamics. For stress imposed dynamics we describe further three different protocols employed to trigger avalanches. We then provide in Sec.~\ref{sec:avalanche:stat} a first comparison of the distributions of avalanche sizes measured numerically for different protocols. 
In Sec.~\ref{sec:control:stat}
we examine the impact of the avalanche-triggering mechanism and of the choice of control parameter on avalanche size distributions. 
Then, Sec.~\ref{sec:fss:main} focuses on finite-size effects and on the associated scaling relations.
We determine via a finite-size scaling analysis the fractal dimension and dynamical exponents. 
This section also discusses the validity of a hyperscaling relation between exponents for stress-controlled avalanches. 

By systematically studying the impact of different protocols on avalanche dynamics, this work 
emphasizes the importance of the control parameter and the associated protocol choices in determining avalanche statistics, and provides guiding principles on how to interpret avalanche statistics in the many cases where the ``control'' parameter is not a matter of choice (e.g. in earthquake dynamics).

\section{Model and avalanche-triggering protocols}
\label{sec:model:protocols}

\subsection{Elastoplastic model}

To investigate the role of protocols on avalanches distributions in amorphous solids, we perform numerical simulations of an elastoplastic model~\cite{nicolas_deformation_2018}. 
The material is spatially discretized in an ensemble of interacting mesoscopic elements, each of them representing a set of a few constitutive elements (bubbles, drops, particles, etc). 
A mesoscopic element $i$ can be in two different states, elastic ($n_i=0$) or plastic ($n_i=1$). It bears a local strain $\epsilon_i = \epsilon^\mathrm{pl}_i + \epsilon^\mathrm{el}_i$ which is decomposed into a plastic strain $\epsilon^\mathrm{pl}_i$ and an elastic strain $\epsilon^\mathrm{el}_i$, and an elastic stress $\sigma_i = \mu \epsilon^\mathrm{el}_i$. 
When in an elastic state, the plastic strain remains constant and therefore the accumulated strain is purely elastic.
Once the local stress overcomes some local threshold $\sigma_\mathrm{Y}$, an element can turn plastic, according to some model-specific rules.
When in a plastic state, %the local strain is constant, 
the local stress relaxes, and is redistributed to other elements in the material via elastic interactions. 
In turn, plastic elements switch back to the elastic state, again according to some model-specific rule.
An elastoplastic model is then defined by the rules coupling the evolution of the three local variables $(n_i, \sigma_i, \epsilon_{pl,i})$.
Elastoplastic models show avalanches because of the stress redistribution occuring when an element is plastic, that can in turn induce plasticity in other elements. 

To demonstrate the dependence of the different triggering protocols, we choose to implement the twodimensional Picard model \cite{picard_slow_2005}, one of the simplest elasto-plastic models in the literature. $N=L^2$ elements are positioned on a square lattice with main axis $\bm{e}_x$ and $\bm{e}_y$. 
The stress and strain are scalar, and correspond to the $xy$ components of the stress and strain tensors, respectively.
On each site, they evolve according to
\begin{equation}
    \partial_t\sigma_i = \sum_{j}G_{ij}\partial_t\epsilon_{j}^{\mathrm{pl}}+\mu\dot{\gamma}, \quad
    \partial_t\epsilon_{i}^\mathrm{pl} = n_i\sigma_i 
    \label{eq:Picard}
\end{equation}
with $\dot{\gamma}$ the global strain rate. The kernel $G_{ij}$ is the discretized version of the Eshelby kernel $\mathcal{G}(\bm{r} - \bm{r}^\prime)$ \cite{eshelbyDeterminationElasticField1957,picard_elastic_2004} which represents the far-field effect of a plastic deformation located in $\bm{r}^\prime$ on the stress in position $\bm{r}$ in a linear elastic medium 
\begin{equation}
    \mathcal{G}(\bm{r} - \bm{r}^\prime) = \frac{\cos(4\theta)}{\pi \lvert \bm{r} - \bm{r}^\prime \rvert^2 }
\end{equation}
with $(\bm{r} - \bm{r}^\prime) \cdot \bm{e}_x = \lvert \bm{r} - \bm{r}^\prime \rvert \cos \theta$.
Separately, transitions between elastic and plastic states are governed by
\begin{equation}
    \left\{
    \begin{array}{lcc}
    n_i: & 0\xrightarrow{\tau}1 & |\sigma_i|>\sigma_\mathrm{Y} \\
    n_i: & 0\xleftarrow{\tau}1 & \forall \sigma_i\\
    \end{array}
    \right.
    \label{eq:Picard_rates}
\end{equation}
with $\tau^{-1}$ the probability rate.

In practice, the time evolution is computed via an Euler method and the convolution is performed using a FFT-based algorithm, in which the discretized kernel is defined as
\begin{equation}
    \hat{G}_{\bm{q}} = -4\frac{q_x^2q_y^2}{q^4}\, ,
\end{equation}
for $\bm{q}\neq 0$ and $\hat{G}_0=-1$.

In its original version, the Picard model was designed to perform simulations at a constant strain rate $\dot{\gamma}$. In this case, the term $\mu\dot{\gamma}$ in \autoref{eq:Picard} is constant. It is nevertheless possible to switch the control parameter, setting the global stress $\Sigma=\frac{1}{N}\sum_i\sigma_i$ as a constant and letting $\dot\gamma$ evolve accordingly~\cite{liu_creep_2018}, following $\dot{\gamma}=\frac{1}{N}\sum_i\partial_t\epsilon_{i}^\mathrm{pl}$. This model then enables us to access the macroscopic behavior of the system in imposed strain or imposed stress dynamics.

This model recovers the yield-stress fluid phenomenology \cite{nicolas_deformation_2018}.
When the strain rate $\dot\gamma$ is controlled, steady-state flow follows a Hershel-Bulkley flow curve $\Sigma = \Sigma_\mathrm{Y} + K \dot\gamma^n$, with $K$ and $n$ constants and $\Sigma_Y$ the macroscopic yield stress, as shown in \autoref{fig:CritDyn}.
When the stress $\Sigma$ is controlled, steady-state flow is only possible for $\Sigma > \Sigma_\mathrm{Y}$, in which case it follows the same flow curve. 
For $\Sigma < \Sigma_\mathrm{Y}$ the deformation stops after a finite strain and the system remains in an elastic mechanical equilibrium.

In this work, we restrict ourselves to the case $\Sigma > 0$ and deal with avalanches under controlled stress at $\Sigma = \Sigma_\mathrm{Y}$. In order to determine precisely $\Sigma_\mathrm{Y}$, we perform simulations at a constant strain rate to build the flow curve $\dot{\gamma}=f(\Sigma)$. The yield stress $\Sigma_\mathrm{Y}$ is determined as the value for which $\log \dot{\gamma}$ is a linear function of $\log[\Sigma-\Sigma_\mathrm{Y}]$ for small $\dot\gamma$, which corresponds to the algebraic scaling $\dot\gamma \sim (\Sigma - \Sigma_\mathrm{Y})^{1/n}$.

\subsection{Avalanches in strain-controlled protocols}

\begin{figure}[h]
    \centering
    \includegraphics[width=0.4\textwidth]{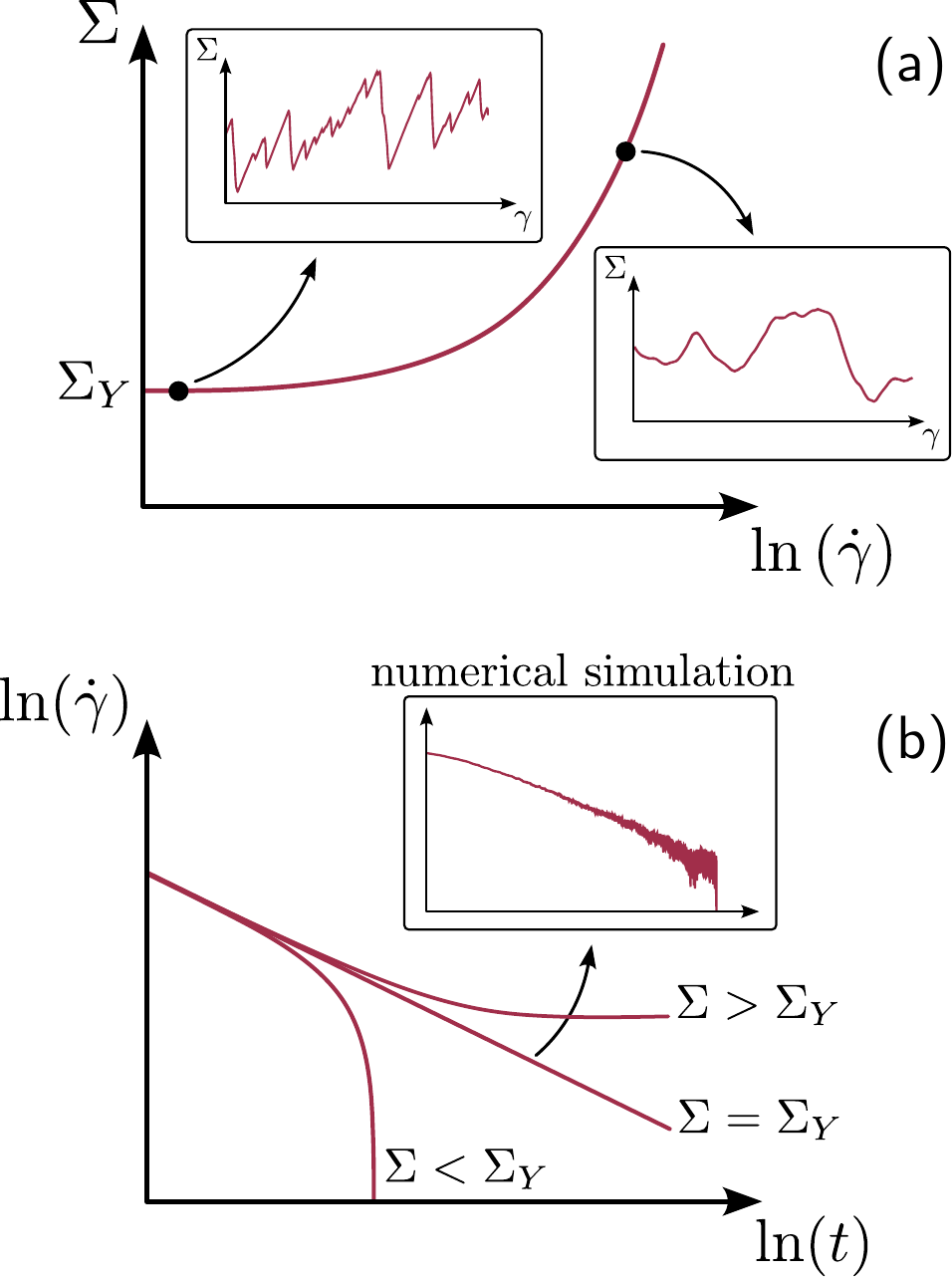}
    \caption{(a) Flow curve of the model, showing a Hershel-Bulkley behavior. In insets, actual numerical simulations of the dynamics under controlled strain rate $\dot\gamma$, at small and large values of $\dot\gamma$, revealing a jerky dynamics with avalanches for $\dot\gamma \to 0$. (b) Schematic expected strain rate versus time under imposed constant stress $\Sigma$ around the yield value $\Sigma_\mathrm{Y}$, showing either (i) arrest of the dynamics for $\Sigma <\Sigma_\mathrm{Y}$, (ii) eventual steady flowing for $\Sigma > \Sigma_\mathrm{Y}$, and (iii) power-law decay for $\Sigma =\Sigma_\mathrm{Y}$. In inset, actual numerical simulations at $\Sigma =\Sigma_\mathrm{Y}$, showing that in practice the power-law decay is interrupted after a finite time at which the dynamics stops in an elastic state.}
    \label{fig:CritDyn}
\end{figure}

In a strain-controlled protocol, as $\dot{\gamma}\rightarrow 0$ , the dynamics becomes jerky. 
The system jumps regularly from one elastic state (i.e., a state where all elements are elastic) to the next via bursts of plasticity. 
Looking at the evolution of the global stress as a function of the global strain, this corresponds to a succession of elastic loading phases and sudden drops of the global stress (see \autoref{fig:CritDyn}). 
These bursts of plasticity in the system are strain-controlled avalanches.

The standard protocol to study this phenomenon is the athermal quasi-static protocol (AQS), described in~\autoref{fig:protocols}(a), which corresponds to the dynamics in \autoref{eq:Picard} and \autoref{eq:Picard_rates} in the $\dot\gamma \to 0$ limit. 
Starting from an elastic state, the system is strained until a site reaches its yield value. 
This site then turns plastic, the straining process is stopped and plasticity propagates in the system at $\dot{\gamma}=0$ such that, through an avalanche process, the system falls back into a new elastic state. 
The system is then strained again until a new avalanche gets triggered and so on. The process is repeated to get statistics about those relaxation events.

Avalanches are mainly characterized by their size $S$ and their duration $T$.  In a system of size $L$, the size of an avalanche is defined as the global stress drop due to the plastic activity multiplied by the volume of the system $S_\Sigma=\Delta\Sigma\times L^d$. The duration of the avalanche simply corresponds to the time between the start and the end of it. In an infinite size system those two quantities are expected to show probability density functions $P_{S_\Sigma}(S_\Sigma)$ and $P_T(T)$ decaying as power laws. These power-law scalings define the avalanche exponents $\tau$ and $\tau^\prime$ as $P_{S_\Sigma}(S_\Sigma)\sim S_\Sigma^{-\tau}$ and $P_T(T)\sim T^{-\tau^\prime}$.
However, in a finite-size system of size $L$, these power laws are only valid up to some cut-off $S_c$ and $T_c$ scaling algebraically with $L$ as $S_c\sim L^{d_f}$ and $T_c \sim L^z$, which defines the fractal dimension $d_f$ and the dynamical exponent $z$. 
Distributions of avalanche size and of avalanche duration for all system sizes $L$ are expected to follow scaling behaviors
$L^{\tau d_f} P_{S_\Sigma}(S_\Sigma) = f_{S_\Sigma}(S_\Sigma L^{-d_f})$ and $L^{\tau^\prime z} P_T(T) = f_T(T L^{-z})$.

\subsection{Avalanches in stress-controlled protocols}

If one imposes a stress $\Sigma$ such that $\Sigma \leq \Sigma_\mathrm{Y}$, the system stops flowing after a finite strain and get stuck in an elastic equilibrium that acts like an absorbing state~\cite{jocteurYieldingAbsorbingPhase2024a}, effectively preventing any measurement of the avalanche statistics by lack of proper sampling.
For finite-size systems, this problem is exacerbated, as due to fluctuations the dynamics can be stuck into an elastic state even for $\Sigma \gtrsim \Sigma_Y$ (see \autoref{fig:CritDyn}).

\begin{figure}
    \centering
    \includegraphics[width=0.4\textwidth]{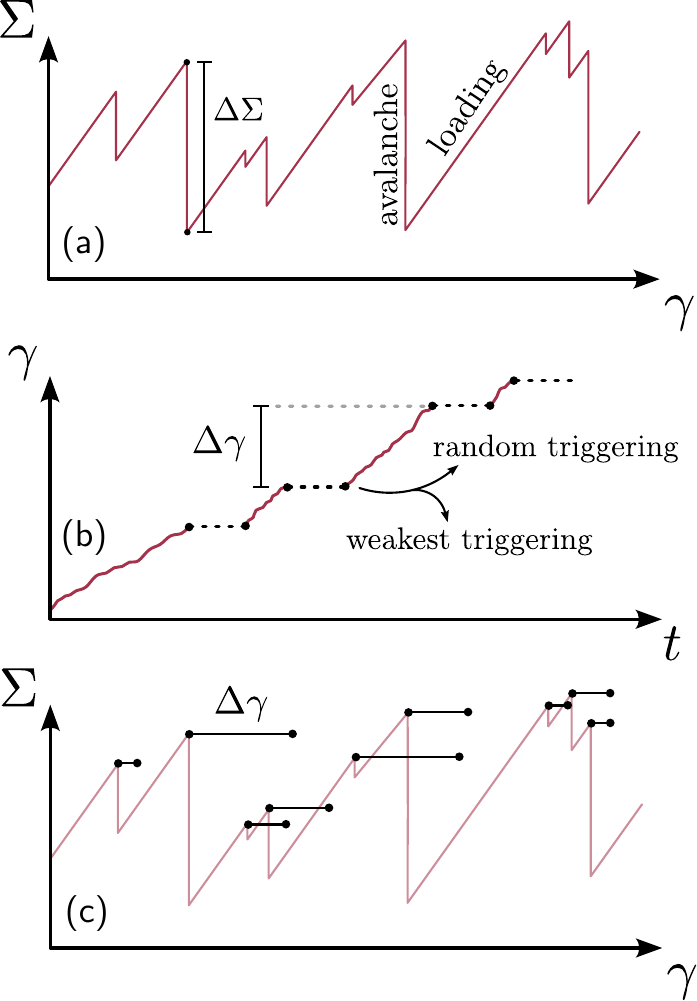}
    \caption{Graphical representation of the different protocols studied. (a) Stress versus strain in the strain-controlled Athermal quasi-static (AQS) dynamics. (b) Strain versus time in the stress-controlled dynamics used in weakest triggering and random triggering protocols. (c) Stress versus strain in the uniform loading protocol.}
    \label{fig:protocols}
\end{figure}

In order to probe the dynamical behavior and sample the avalanche statistics near the yield point, one has to reactivate the dynamics every time the system gets stuck in an elastic state~\cite{lin_scaling_2014}. 
An avalanche is then defined as the plastic activity between two successive reactivations, as sketched in~\autoref{fig:protocols}(b). 
The avalanche size is the accumulated strain during the avalanche multiplied by the volume of the system, $S_\gamma=\Delta\gamma\times L^d$.

There are several ways to reactivate the system, defining different protocols to generate stress-controlled avalanches. 
It is then legitimate to wonder whether this arbitrary choice of protocol may have an impact on the produced avalanches and the statistical observables that characterize them. 

\subsection{Definition of the protocols}

We study two different stress-controlled protocols, defined by the reactivation mechanism: the random triggering protocol, and the weakest triggering protocol. 
Both of them are studied at the critical stress $\Sigma = \Sigma_\mathrm{Y}$.
The avalanche statistics produced by these two protocols is studied in the steady-state, after many reactivations are performed.
As shown below, each reactivation mechanism gives a different statistics of visited elastic states.
To disentangle the effect of controlling the stress during the avalanche from the effect of exploring statistically different elastic states, we also study a last protocol, the uniform loading protocol, which uses starting states sampled from AQS dynamics, but performs avalanches under imposed stress.

\subsubsection{Random triggering protocol}

The first stress-controlled protocol we studied is the random triggering protocol. 
Every time the system falls into an elastic state, the system is reactivated by setting a random site, chosen with uniform probability, as plastic ($n_i=1$) without changing the stress anywhere in the system. 
Experimentally, this protocol could be realized with local damage in the sample, as is performed in~\cite{dallari_stochastic_nodate}. 
Note that this protocol is similar to the one used in the numerical study reported in~\cite{lin_scaling_2014}. The only difference is that
in our case activation is made by a direct change on the state variable $n_i$, while in~\cite{lin_scaling_2014} it is made by giving randomly localized stress kicks to the system.

\subsubsection{Weakest triggering protocol}

The second stress-controlled protocol is the weakest triggering one. 
In this protocol, a single site $i$ ($n_i=1$) is activated as in the random triggering protocol, but now the activated site is the one closest to local yield, i.e., with the smallest value of $\sigma_\mathrm{Y} - \sigma_i$. 
Here again, it resembles one of the numerical protocols proposed in~\cite{lin_scaling_2014} where the weakest site would be activated, only not by setting $n_i=1$ but by giving it a stress kick.

\subsubsection{Uniform loading protocol}

In the last protocol, denoted as uniform loading protocol, the dynamics is not triggered from the elastic state reached after the previous avalanche.
We first generate a long AQS simulation that contains many avalanches.
We then record the starting configuration of each AQS avalanche, which corresponds to the point where one site (i.e., the marginal site) is at
$\sigma_i = \sigma_\mathrm{Y}$.
Among these, we select the ones for which $\Sigma < \Sigma_\mathrm{Y}$.
For each of these configurations, the avalanche starts from the marginal site and evolves under a constant stress dynamics, as sketched in \autoref{fig:protocols}(c), until the system reaches again a purely elastic state, which is guaranteed by our choice of states with $\Sigma < \Sigma_\mathrm{Y}$.

Because they come from an AQS dynamics, starting configurations are all at a slightly different stress below $\Sigma_\mathrm{Y}$, nonetheless the stress is always close to $\Sigma_\mathrm{Y}$ (for $L=512$, we get $\Sigma\gtrsim 0.995~\Sigma_Y$).
This protocol differs from the one studied in~\cite{budrikis_universal_2017} as the elastic states we reach after an avalanche are not used again to trigger further avalanches, in contrast to~\cite{budrikis_universal_2017} where the system is elastically loaded to trigger the next avalanche, leading to states of strictly increasing stresses, and eventually to runaway avalanches and permanent flow.

\begin{figure}
    \centering
    \includegraphics[width=\columnwidth]{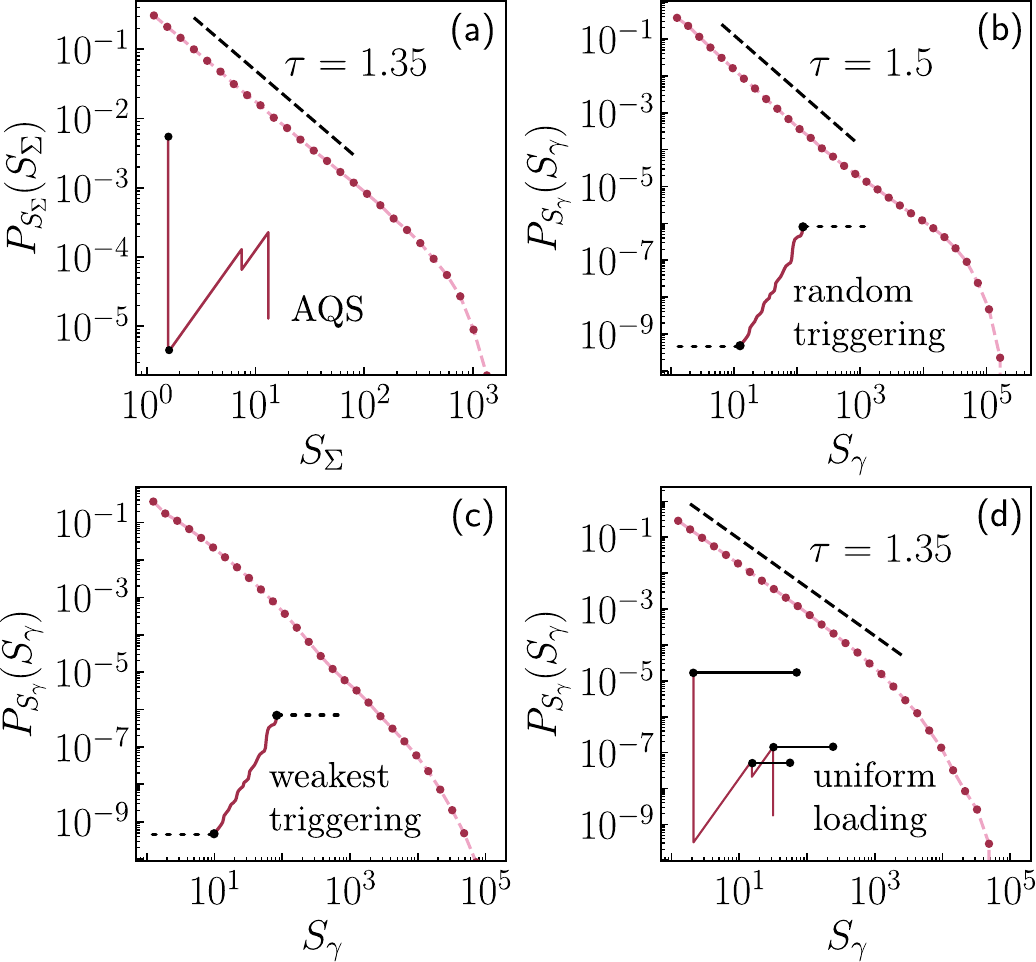}
    \caption{Size distributions of the avalanches for a system size $L=512$ for the four different protocols. (a) AQS protocol. (b) random triggering protocol. (c) weakest triggering protocol. (d) uniform loading protocol.}
    \label{fig:Distributions}
\end{figure}

\section{Avalanche statistics for different protocols}
\label{sec:avalanche:stat}

For each of our four protocols (AQS, random triggering, weakest triggering and uniform loading), we generate about $5\times10^5$ avalanches to build the avalanche statistics. Results are shown in~\autoref{fig:Distributions}.

As expected, for the AQS protocol (\autoref{fig:Distributions}(a)) we observe power-law-distributed event sizes with an avalanche exponent $\tau\approx1.35$,
in agreement with several previous studies of the AQS protocol in different elastoplastic models~\cite{ferrero_criticality_2019, liu_driving_2016, lin_scaling_2014}. 
The power law is observed up to some cut-off $S_\mathrm{c}(L)$ past which the distribution roughly decays exponentially.

The uniform loading protocol (\autoref{fig:Distributions}(d)) produces a very similar size distribution to the one of the AQS protocol as it is also power-law distributed with an exponent $\tau\approx 1.35$. 

Our analysis suggests that the control parameter, whether strain or stress, exerts a minimal impact on avalanche properties. This influence is notably less pronounced than that of the nature of the activated elastic states, indicating a secondary role for the control parameter in modulating avalanche dynamics.
%This may suggest that the control parameter (strain or stress) \emph{per se} has a limited impact on avalanches properties.
% , or at least much less than the nature of the activated elastic states. 
Nevertheless, the tail of the distribution significantly differs from that obtained with the AQS protocol. 
The cut-off is significantly larger than the one of the AQS protocol, similarly to the ones obtained in the other stress-controlled protocols, as shown below.

In the random triggering protocol (\autoref{fig:Distributions}(b)), avalanche sizes are still distributed according to a clear power law but with a significantly different exponent $\tau\approx1.5$, in agreement with the theoretical mean-field prediction $\tau=3/2$~\cite{dahmen_simple_2011}. This variation underscores the sensitivity of avalanche dynamics to the chosen protocol, highlighting the importance of the initial triggering mechanism. Note that this result contradicts earlier numerical observations~\cite{lin_scaling_2014}, for which the random triggering protocol was reported to give the same exponent $\tau$ as the AQS dynamics, smaller than our measured value.

Just like for uniform loading, the cut-off value is larger than the one found for the AQS protocol by roughly two orders of magnitude. 
Furthermore, the shape of the distribution deviates from a pure power law for large avalanche sizes, the cut-off being preceded by a characteristic bump showing an excess of avalanches with respect to the power-law expectation. 
Such bumps have been observed for other protocols in the literature~\cite{budrikis_universal_2017, sandfeld_avalanches_2015, oyama_unified_2021}, and are predicted by the renormalization group analysis of depinning models~\cite{rosso_avalanche-size_2009}. 
This bump may affect the measure of the $\tau$ exponent, leading to the determination of an effective exponent smaller than the actual one, as argued in~\cite{oyama_unified_2021}. 
As shown below, finite-size scaling alleviates much of this difficulty, and considerably narrows the uncertainty of our reported value of $\tau$.

Finally, the weakest triggering protocol (\autoref{fig:Distributions}(c)) gives vastly different avalanche statistics. 
Avalanche sizes are broadly distributed but not according to a power law. 
Instead, the distribution shows two successive bumps, as also visible in the data of~\cite{lin_scaling_2014} (Fig. S4 therein). 
The cut-off for similar system sizes is however of the same order of magnitude as the one of the random triggering protocol. 

These results  demonstrate that the reactivation protocol when studying avalanches under controlled stress has a major impact on the measured avalanche statistics. 
Avalanche statistics under controlled stress is therefore not strictly a system property, and one needs to be explicit as to which protocol is used to generate avalanches.
In the following, we try to infer broad relations between generic features of reactivation protocols and resulting avalanche statistics.

\section{Protocol features that control the avalanches size statistics}
\label{sec:control:stat}

\subsection{The triggering mechanism controls the avalanche exponent $\tau$}

The AQS and uniform loading protocols have the same measured value for the avalanche exponent, $\tau\approx 1.35$.
These two protocols differ by their control parameter, but share the same triggering mechanism, which suggest that the latter controls the value of $\tau$.
Indeed, changing the triggering mechanism affects the exponent value, as the random triggering protocol yields $\tau\approx 1.5$ and the weakest triggering protocol shows no exponent at all.

This is consistent with mean-field arguments which suggest that the exponent characterizing the size-distribution of avalanches is controlled by the way elastic configurations are activated, either by the activation of one random site (as the random triggering protocol) or the uniform loading of all sites (as the AQS protocol)~\cite{jagla_avalanche-size_2015}. 
In these cases, the mean-field problem can be mapped onto a random walk first-passage problem and yield exact analytical results \cite{jagla_avalanche-size_2015}. 
For random triggering, an exponent $\tau = 1.5$ is predicted.
By contrast, for uniform loading, the mean-field prediction is an effective exponent which slowly varies as a function of avalanche size. 
For typical sizes simulated here, an exponent of $\tau\approx 1.1$ is predicted. 
Our results show that the qualitative mean-field expectation of a smaller (effective) exponent  when avalanches are triggered by elastic loading rather than random activation 
still holds in finite dimension. 

However, quantitative predictions of the mean-field scenario \cite{jagla_avalanche-size_2015} do not match several of our numerical observations.
First, the mean-field prediction $\tau\approx 1.1$ for the uniform loading protocol is significantly below our measured value $\tau\approx 1.35$.
Second, the mean-field calculations predict a bump in the distribution for large avalanches in the case of the uniform loading, and no bump for random activation, which is the opposite of our observations.

Overall, these results suggest that the avalanche exponent is not controlled by the dynamical rules that apply during avalanches but rather that it strongly depends on the way avalanches are initiated. 
Loaded elastic states show smaller exponents than locally triggered ones. 
However, if dynamical rules do not seem to affect the exponent, they still have an impact on the avalanche distribution. Looking at \autoref{fig:Distributions}, it is clear that all stress-controlled protocols show much larger cut-offs than the one of the AQS protocol.

\subsection{The control parameter controls the cut-off size}
\label{sec:control_cutoff}

A major difference between stress-controlled and strain-controlled dynamics is that during a strain-controlled avalanche the total stress drops, effectively making the avalanche more likely to stop early. 
Thus, starting from the same state, we would expect the avalanche in a stress-controlled protocol to be larger and last longer
than in a strain-controlled protocol.
This is indeed confirmed by our observations in~\autoref{fig:Distributions}, which shows that the cutoff size is smaller for the AQS protocol.

\begin{figure}
    \centering
    \includegraphics[width=\columnwidth]{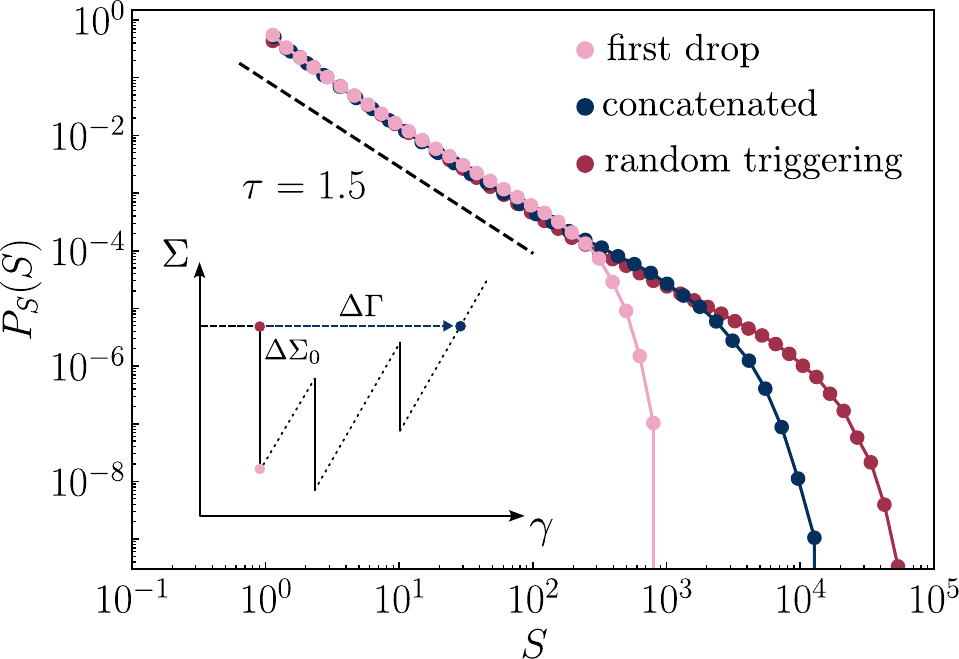}
    \caption{Size distributions of avalanches in the concatenation procedure for a system size $L=256$. Inset: schematic of the concatenation procedure. First drop corresponds to the first strain-controlled avalanche (between the first red dot and the blue dot in the schematic). Concatenated avalanches correspond to the sum of AQS avalanches required to recover the initial global stress $\Sigma_Y$.}
    \label{fig:Concatenation}
\end{figure}

Under AQS dynamics, if the starting point of an avalanche is at a relatively high stress, it may take several successive avalanches and elastic loadings to find again an elastic state at the same stress. 
It is thus tempting to sketch avalanches under controlled stress as concatenations of successive AQS-like avalanches that occur between two states whose stress matches the imposed one, covering a strain $\Delta\Gamma$, as illustrated in the inset of \autoref{fig:Concatenation}.

To investigate this idea, we run an AQS dynamics from the starting configurations of every avalanche obtained with the random triggering protocol.
These configurations are all initially at $\Sigma = \Sigma_\mathrm{Y}$ by construction. 
During the AQS dynamics, their stress drops during avalanches and increases during elastic loading phases, until it reaches $\Sigma_\mathrm{Y}$ again, at which point we stop the simulation and record the strain $\Delta\Gamma$ since the start of the AQS dynamics.

In \autoref{fig:Concatenation}, we display in dark blue the obtained distribution of $S_{\Gamma} = \Delta\Gamma L^d$, alongside the distribution of avalanche sizes measured in the original random triggering protocol.
To better highlight the effect of the concatenation on the statistics of avalanches, we also show the distribution of the first stress drop size $P_{S_{\Sigma_0}}(S_{\Sigma_0})$, with $S_{\Sigma_0} = \Delta\Sigma_0 L^d$, during the AQS dynamics (pink curve): 
if each avalanche in the random triggering protocol corresponded to a single AQS avalanche followed by a single elastic loading to get back to a stress $\Sigma_\mathrm{Y}$, $P_{S_\gamma}(S_\gamma)$ would be indistinguishable from $P_{S_{\Sigma_0}}(S_{\Sigma_0})$ (and $P_{S_\Gamma}(S_\Gamma)$ for that matter). 
We see that this is the case for small avalanches (we checked that this is not only true statistically, but that pairs of small avalanches under random triggering and AQS sharing the same starting point are indeed very similar).
However, large stress-controlled avalanches are clearly not associated with single AQS avalanches.

The distribution of intermediate size avalanches under random triggering is better approximated by the distribution of concatenated AQS avalanches, 
showing that some large stress-controlled avalanches seem to be reasonably well approximated by concatenation of AQS avalanches.
However, there remains a clear discrepancy between $P_{S_\gamma}(S_\gamma)$ and $P_{S_\Gamma}(S_\Gamma)$ for the largest avalanches, that interestingly correspond to avalanches in the already noted excess bump.

Massive avalanches under stress-controlled conditions are therefore more than bare successions of AQS-like avalanches. 
This indicates that the finite strain rate during stress-controlled avalanches can no longer be neglected for these avalanches.
Large avalanches transiently lead to large strain rates in the system, 
moving the system trajectory away from the one followed by the AQS dynamics.
In particular, under finite strain rates sites become plastic at stresses larger than $\sigma_{\mathrm{Y}}$ as elastic loading persists before the stochastic triggering of plasticity, causing plastic sites to redistribute larger stresses than they would in a similar configuration under AQS.
This larger redistribution in turn fuels more plastic triggering, eventually leading to larger avalanches, which is compatible with our observation of a larger cutoff value for $P_{S_\gamma}(S_\gamma)$ than for $P_{S_\Gamma}(S_\Gamma)$ in \autoref{fig:Concatenation}.

\section{Finite-size scaling and scaling relations}
\label{sec:fss:main}

\subsubsection{Fractal dimension and dynamical exponents}

We now turn to the fractal dimension $d_f$ of avalanches and dynamical exponent $z$, which we evaluate using a finite-size scaling analysis.
We first confirm the results reported in the literature for the AQS protocol, that is $L^{\tau d_f} P_{S_\Sigma}(S_\Sigma)$ plotted as a function of $S_\Sigma L^{-d_f}$ and $L^{\tau^\prime z} P_T(T)$ plotted as a function of $T L^{-z}$ respectively show a data collapse onto a master curve for $d_f\approx 1.1$ and $z\approx 0.54$~\cite{liu_driving_2016} (see \autoref{fig:finite-size-scaling-AQS} in Appendix~\ref{sec:fss:app}).

\begin{figure}
    \centering
    \includegraphics[width=\columnwidth]{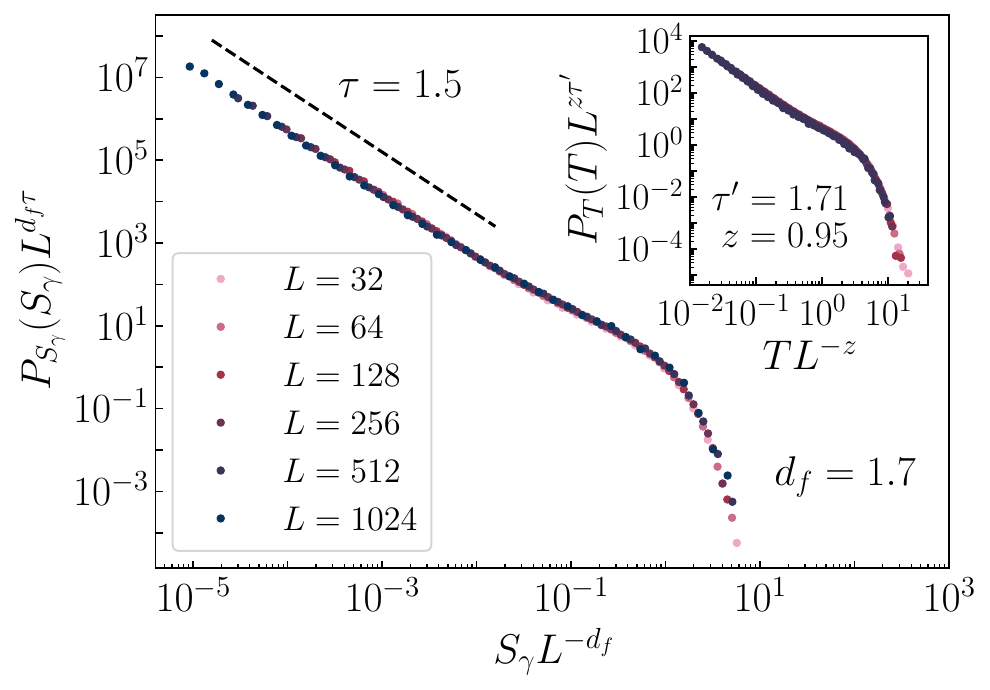}
    \caption{Finite-size scaling of the avalanche size distributions in the RSA protocol for system sizes $L\in [32, 64, 128, 256, 512, 1024]$. The best collapse determines the best estimate for the fractal dimension $d_f\approx1.7$. Inset: Finite-size scaling of the avalanche duration distributions in the random triggering protocol for the same system sizes. The best collapse determines the best estimate for the dynamical exponent $z\approx0.95$.}
    \label{fig:finite-size-scaling}
\end{figure}

Remarkably, we can achieve a similar collapse for avalanches obtained with the random triggering protocol, as displayed in \autoref{fig:finite-size-scaling}. 
Using distributions for system sizes ranging from $L=32$ to $L=1024$, the best collapse of the curves is obtained for $\tau\approx1.5$, $d_f\approx1.7$, $\tau^\prime\approx1.71$ and $z\approx0.95$. The fractal dimension obtained in this case is then much larger than the one measured for the AQS protocol. 

This is consistent with the picture that, at least for intermediate-sized events, stress-controlled avalanches are close to a concatenation of AQS-like avalanches. 
Indeed, successive AQS avalanches are known to be spatially correlated and to cluster along the flow and flow gradient directions~\cite{martensConnectingDiffusionDynamical2011}.
To confirm this hypothesis,  we perform the same finite-size scaling for the distributions of concatenated AQS avalanches, and also find $d_f\approx 1.5$ (see \autoref{fig:finite-size-scaling-concatenated} in Appendix~\ref{sec:fss:app}), in contrast to the $d_f \approx 1.1$ for individual AQS avalanches (which we also observe for the distribution of first stress drops $P_{\Delta\Sigma_0}(\Delta\Sigma_0)$).

Finally, despite the fact that for the weakest triggering protocol the distributions of avalanche size and duration do not show well defined power laws, their cutoff still scale as a power law of the system size, as shown in \autoref{fig:finite-size-scaling-weakest}. This allows us to determine exponents $d_f\approx 1.4$ and $z\approx 0.75$ for this protocol, which lie in between the values obtained for AQS and those obtained with random triggering.

\subsubsection{Relation to the pseudo-gap exponent}

Scaling relations relating the avalanche behavior of amorphous systems to their structure or flowing properties have been proposed in the past and seem to be satisfied in the case of the AQS protocol \cite{lin_scaling_2014, liu_driving_2016}. 
One of them relates exponents $\tau$ and $d_f$ with the pseudo-gap exponent in the distribution of distance to the local yield $\sigma_\mathrm{Y}$. 
Defining for each site the distance to local yield $x_i = \sigma_\mathrm{Y} - \sigma_i$, for $\Sigma = \Sigma_\mathrm{Y}$ one expects at small $x$ a power-law behavior for the distribution $P_x(x) \sim x^\theta$, with $\theta$ the pseudo-gap exponent \cite{karmakarStatisticalPhysicsYielding2010,linDensityShearTransformations2014}. 

\begin{figure}
    \centering
    \includegraphics[width=0.48\textwidth]{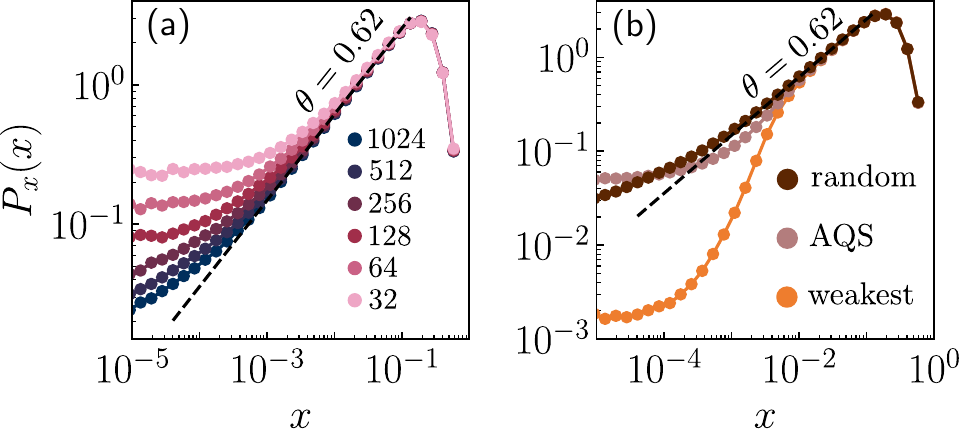}
    \caption{(a) Distributions of local distance to threshold in the elastic states of the random triggering protocols for system sizes $L\in [32, 64, 128, 256, 512, 1024]$. The dotted black line indicates the universal power-law determined with an exponent $\theta\approx 0.62$. (b) Distributions of local distance to threshold in the elastic states of different protocols for a system size $L=512$. The weakest triggering protocol shows a clear depletion at low $x$ compared to the other protocols.}
    \label{fig:P(x)}
\end{figure}

Under AQS protocol, stationarity imposes that \cite{lin_scaling_2014}
\begin{equation}
    \tau = 2 - \frac{\theta}{\theta+1}\frac{d}{d_f}\, .
\end{equation}
One may wonder whether this scaling relation is also valid beyond the AQS protocol, and in particular for the random triggering protocol, which shows a scaling behavior, albeit with different exponent values from the ones observed in AQS. 
We construct the distribution $P_x(x)$ for the random triggering protocol 
by sampling the elastic states obtained between two avalanches. 
Statistics are here obtained for a total of approximately $2\times 10^{10}$ individual $x$ values. 
In \autoref{fig:P(x)}(a), we show $P_x(x)$ for several system sizes. 
For each size, we observe a power law with an exponent $\theta\approx 0.62$ for $x$ in an intermediate range between a small $x$ plateau and a large $x$ peak.
The larger the system size, the broader the power-law range at the lower end, which supports the claim that $\theta\approx 0.62$ is the infinite size limit exponent. 
Using this measurement of $\theta$ along with $d_f$ and $\tau$ measured by finite-size scaling (see above), we find $\tau d_f/[2d_f-\theta d/(\theta+1)] \approx 0.97$, which shows that the scaling relation is still valid to a good accuracy for the random triggering protocol.

The weakest triggering protocol shows no scaling behavior in the avalanche size distribution, and therefore the scaling relation is not applicable in this case. 
Nonetheless, it is interesting to look at the distribution $P_x(x)$ for this protocol, as shown in \autoref{fig:P(x)}(b) and compare it to the one obtained for the AQS and random triggering protocols with the same system size. 
Interestingly we find that there still is an intermediate range of $x$ for which we observe the pseudo-gap behavior, with the same $\theta\approx 0.62$, but this range is much narrower for weakest triggering than for AQS or random triggering.
For small $x$ values, the distribution is severely depleted in the case of weakest triggering.
The size dependence of $P_x(x)$ reveals that the pseudo-gap behavior widens with larger system sizes (see Appendix~\ref{sec:p_x_weakest}), but the size dependence is comparatively weak with respect to AQS and random triggering.
This slow emergence of a scaling behavior of $P_x(x)$ suggests that the absence of power law in the avalanche distribution might be a finite-size effect that persists up to unusually large system sizes.
Unfortunately, we cannot explore significantly larger system sizes in order to clarify if the weakest triggering protocol actually follows a scaling behavior in the large system size limit.

\begin{table*}
\caption{Measured exponents in studied protocols. \O: no convincing exponent determined; ---: not measured; ${}^\ast$: incomplete finite-size collapse}
\begin{ruledtabular}
% \begin{tabular}{ccccccccc} 
\begin{tabular}{cccclllll} 
  Protocol & Control variable & elastic states & Activation & $\tau$ & $\tau'$ & $d_f$ & $z$ & $\theta$ \\
 \hline
 AQS & strain & \text{self-generated} & \text{uniform loading} & 1.35 & 1.8 & 1.1 & 0.54 & 0.62 \\ 
 random triggering & stress & \text{self-generated} & $n_i=1$, $i$ random & 1.5 & 1.7 & 1.7 & 0.95 & 0.62 \\
 weakest triggering & stress & \text{self-generated} & $n_i=1$, $i=\text{argmin}\{\sigma\}$ & \O & \O & $1.4^\ast$ & $0.75^\ast$ & 0.62 \\
 uniform loading & stress & AQS & uniform loading & 1.35 & ---& ---& ---& 0.62
 \label{table:exponents}
\end{tabular}
\end{ruledtabular}
\end{table*}

\subsubsection{Relation to the flow exponent}

An additional scaling relation can be derived between the yielding critical exponents and the avalanche exponents $z$ and $d_f$. 
In the flowing phase for $\dot\gamma>0$ small, the strain rate $\dot{\gamma}$ and stress are related via $\dot{\gamma}\sim(\Sigma-\Sigma_Y)^\beta$ with $\beta=1/n$, and the correlation length scales as $\xi\sim(\Sigma-\Sigma_Y)^{-\nu}$. 
Based on the idea that in the critical regime the mean strain rate is controlled by avalanches of typical extension $\xi$, characterized by their size $S_\xi \sim \xi^{d_f}$ and their duration $T_\xi \sim \xi^{z}$, we get
\begin{equation}
    \dot{\gamma}\sim \frac{S_\xi}{T_\xi \xi^d}\sim \xi^{d_f-z-d}.
\end{equation}
Eliminating the distance $\Sigma-\Sigma_Y$ to the yielding point in the definition of $\beta$ and $\nu$ we can get the following scaling relation
\begin{equation} \label{eq:scal:rel:beta}
    \beta = \nu (d-d_f+z)\, .
\end{equation}
This scaling relation has been verified earlier in the case of the AQS protocol~\cite{lin_scaling_2014}. 
To test \autoref{eq:scal:rel:beta} on the random triggering protocol, we use the measured values of $d_f$ and $z$. 
On the other hand, $\nu$ has recently been measured in this model, revealing $\nu=1.13$~\cite{jocteurYieldingAbsorbingPhase2024a}. 
The exponent $\beta$ predicted by the scaling relation \autoref{eq:scal:rel:beta} is then $\beta \approx 1.41$, which should be compared to the measured value of $\beta\approx 1.5$ in this model. 
For the AQS protocol on the other hand we get through the same procedure a predicted exponent of $\beta\approx 1.63$. 
Thus both protocols are in relatively good agreement with the scaling prediction even though the strain-controlled protocol tends to overestimate it while the stress-controlled protocol tends to underestimate it. 

\section{Conclusion}

In this study, we examine the protocol dependence of avalanche dynamics in an elastoplastic model, for protocols which are commonly used in the experimental and numerical literature. 
Our results highlight significant differences in avalanche statistics based on the statistical properties of the elastic states prior to the stress drop events, on the external driving protocol and on the specific protocols employed to trigger local plastic events.
A short overview of these results is compiled in \autoref{table:exponents}.

Consistently with former results in the literature, we find that the athermal quasi-static (AQS) protocol yields avalanches following a power-law distribution with an exponent $\tau\approx1.35$. 
For stress imposed dynamics and a uniform loading protocol our model produces similar power-law distributed avalanche sizes with $\tau\approx1.35$. The tail of the distribution shows a larger cut-off compared to AQS, indicating broader distribution with avalanches of larger sizes.
However, the random triggering protocol yields a different exponent $\tau\approx1.5$, closer to the mean-field prediction $\tau=3/2$. The distribution has a characteristic bump before the cut-off, differing significantly in shape from the AQS protocol.

When triggering the weakest site instead, the size distribution no longer follows a well-defined power law and rather displays two successive bumps. 
This suggests a unique behavior compared to other protocols. 
Yet the distribution of the distance to local yield suggests that a power-law scaling may be recovered, but only for much larger system sizes than for AQS and random triggering.

We showed that the exponent $\tau$ is primarily controlled by the way avalanches are triggered rather than the dynamical rules during avalanches. Protocols with uniform loading tend to display smaller exponents than those with local triggering.

A feature that depends strongly on dynamical rules is the cut-off size of the avalanche distributions. Stress-controlled protocols generally exhibit larger cut-offs compared to strain-controlled protocols. We interpret this phenomenon with an argument based on the concatenation of AQS avalanches.

In agreement with earlier research~\cite{liu_driving_2016, lin_scaling_2014}, our findings confirm that the AQS protocol exhibits a fractal dimension of approximately $1.1$ and a dynamical exponent of about $0.54$. Conversely, the random triggering protocol reveals larger values, with $d_f \approx 1.7$ and $z \approx 0.95$, indicating that stress-controlled avalanches are not only more spatially compact but also persist for longer durations compared to their AQS counterparts.

Despite the change of exponent values from one protocol to the other, we verified that the predicted scaling relation $\tau = 2 - \frac{\theta}{\theta + 1} \frac{d}{d_f}$ still holds for both AQS and random triggering protocols. For the random triggering protocol, the measured exponent $\theta \approx 0.62$ is consistent with the predicted $\tau \approx 1.55$. Also the scaling relation $\beta = \nu (d - d_f + z)$ is reasonably satisfied by our model results, indicating a good agreement with theoretical predictions \cite{lin_scaling_2014}.

The twodimensional model considered here gives results that go beyond former mean-field studies \cite{jagla_avalanche-size_2015}. 
Although we confirm some mean-field features to be valid in finite dimension, 
such as a change of avalanche exponents when going from uniform loading to random triggering, specific features like the appearance of a bump in the distribution crucially depend on the low dimensionality of the model. 
Moreover we are able to describe in detail the complex spatial properties of avalanches that are encoded in the fractal dimension and can be used to test general scaling relations.

We believe our work may help to conceptually better understand the differences which may occur in the measurements of avalanche dynamics depending on the protocol employed. 
Care has to be taken when talking about universality under stress-imposed conditions, where the choice of avalanche-triggering protocol has a significant impact on avalanche statistics. 

Finally, we would like to mention a promising route to test our results in an experimental setup. 
There are practical ways to implement our triggering protocols, even in the case of the random triggering protocol.
It has recently been shown experimentally that it is possible to randomly induce plastic events in a glass using X-ray irradiation \cite{martinelli2023reaching}. This technique may be used to randomly trigger plastic avalanches in a glass under imposed stress conditions, to compare with other driving protocols.

In other systems displaying avalanches it is often possible to either control strain or stress, such as in granular materials \cite{houdoux2021micro} or in simulations of the depinning dynamics of elastic manifolds in disordered environments, either with a force-controlled or a velocity-controlled protocol \cite{rosso_avalanche-size_2009}. 
We believe that our results will have an impact on the study of a large variety of systems displaying avalanche dynamics studied in hard and soft condensed matter physics, materials science, solid mechanics, geology and in seemingly distinct fields like neuroscience, biological systems and active matter.

\begin{figure}[t]
    \centering
    \includegraphics[width=0.8\linewidth]{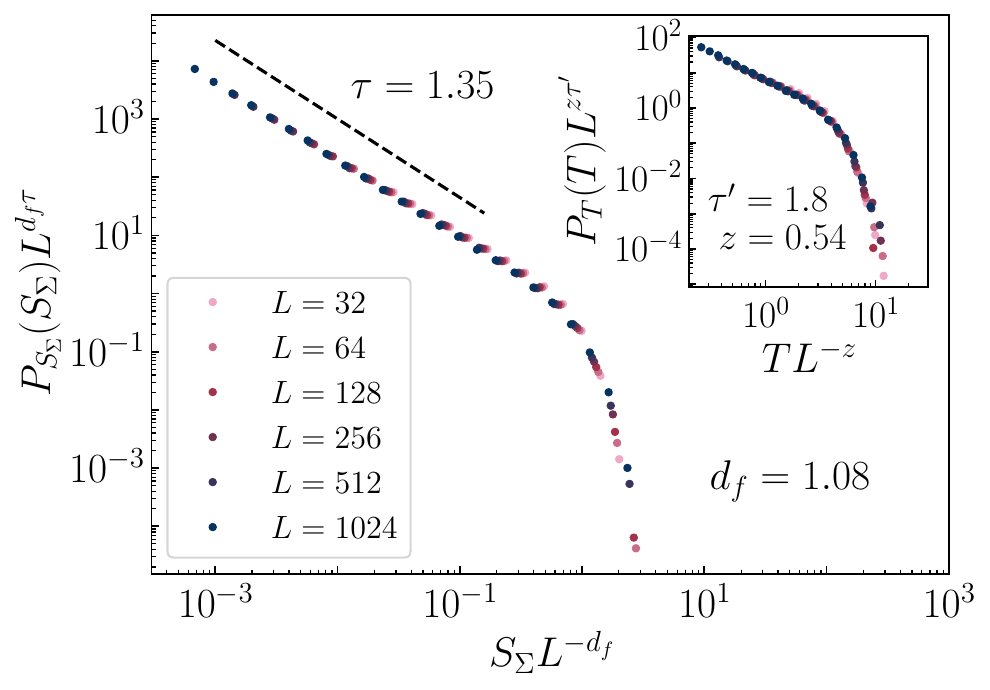}
    \caption{Finite-size scaling of the distribution of AQS avalanche size and duration for several system sizes $L$.}
    \label{fig:finite-size-scaling-AQS}
\end{figure}

\acknowledgments
This project was provided with computer and storage resources by GENCI at
IDRIS thanks to the grant 2023-AD010914551 on the supercomputer Jean Zay's V100 and A100 partitions. 
Some of the computations presented in this paper were performed using the GRICAD infrastructure (\href{https://gricad.univ-grenoble-alpes.fr}{https://gricad.univ-grenoble-alpes.fr}), which is supported by Grenoble research communities.
T.J. acknowledges funding from the French Ministry of Higher Education and Research.

\appendix

\section{Finite-size scaling of avalanches with several protocols} 
\label{sec:fss:app}

\begin{figure}[t]
    \centering
    \includegraphics[width=0.8\linewidth]{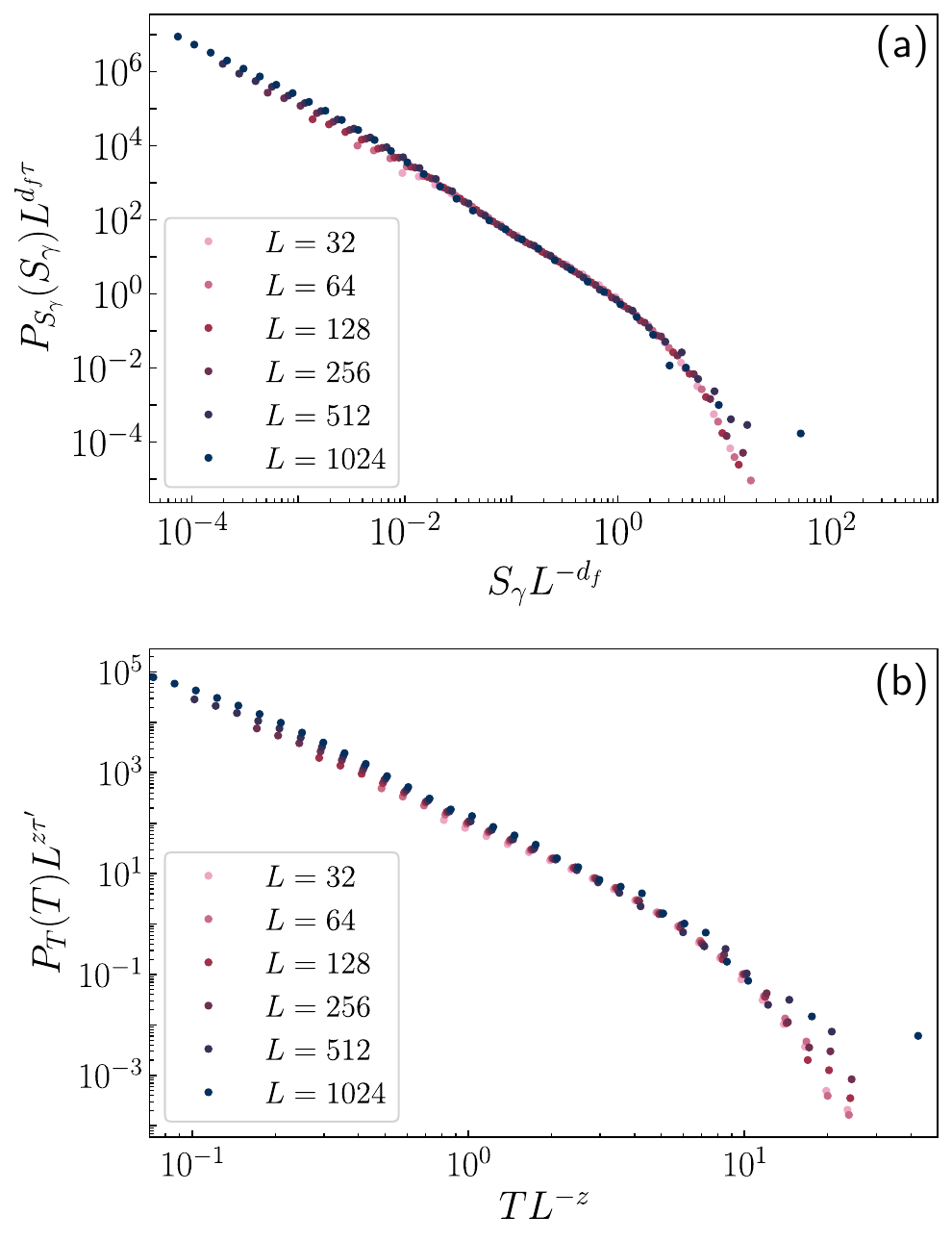}
    \caption{Finite-size scaling of the distribution of avalanche size (a) and duration (b) with the weakest triggering protocol for several system sizes $L$. Trial exponents for this rescaling are $\tau = 1.75$, $d_f = 1.4$, $\tau^\prime = 2.7$ and $z=0.75$.}
    \label{fig:finite-size-scaling-weakest}
\end{figure}

In this appendix we show the results of the finite-size scaling of avalanches in the AQS and weakest triggering protocols.
For AQS, the best collapse of $L^{d_f \tau} P_{S_\Sigma}(S_\Sigma)$ versus $S_\Sigma L^{-d_f}$ is obtained for a fractal dimension $d_f\approx1.08$, using $\tau=1.35$ (see \autoref{fig:finite-size-scaling-AQS}).
The best collapse for $L^{z \tau'} P_T(T)$ versus $T L^{-z}$ is obtained for $z\approx0.54$ and $\tau^\prime \approx 1.8$
(inset of \autoref{fig:finite-size-scaling-AQS}). While the values of $\tau$, $d_f$ and $z$ are close to the ones determined in previous works \cite{liu_driving_2016, lin_scaling_2014}, the value of $\tau^\prime$ is slightly larger than the value 1.4 measured in \cite{liu_driving_2016} and 1.6 measured in \cite{lin_scaling_2014}, which suggests that it could be a model-dependent feature.

For the weakest triggering protocol [\autoref{fig:finite-size-scaling-weakest}], we do not find a satisfying collapse, neither for $P_\Delta\Gamma (\Delta\Gamma)$ nor for $P_T(T)$, as represented in the rescaling trial in $\autoref{fig:finite-size-scaling-weakest}$. This is due to the absence of a clear power law in these distributions, in particular $P_\Delta\Gamma (\Delta\Gamma)$ instead shows two bumps. 
Nonetheless, looking for exponents to collapse best the curves in the cut-off region, we do find a good collapse for large avalanches with $d_f = 1.4$ and $z=0.75$, of course at the cost of having a strong size-dependence for small avalanches.

\begin{figure}[t]
    \centering
    \includegraphics[width=0.8\linewidth]{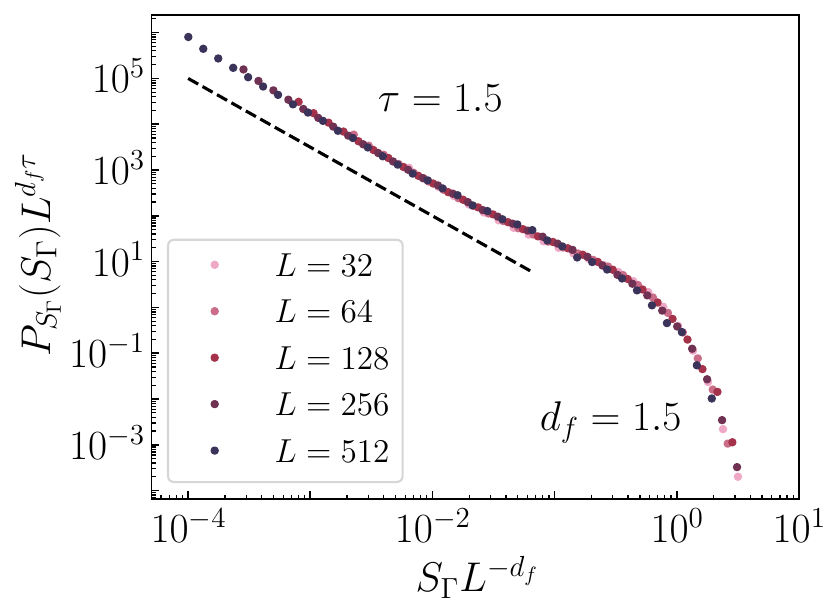}
    \caption{Finite-size scaling of the size distribution of concatenations of avalanches obtained by AQS starting from elastic states sampled by the random triggering protocol for several system sizes $L$.}
    \label{fig:finite-size-scaling-concatenated}
\end{figure}

\begin{figure}[t]
    \centering
    \includegraphics[width=0.8\linewidth]{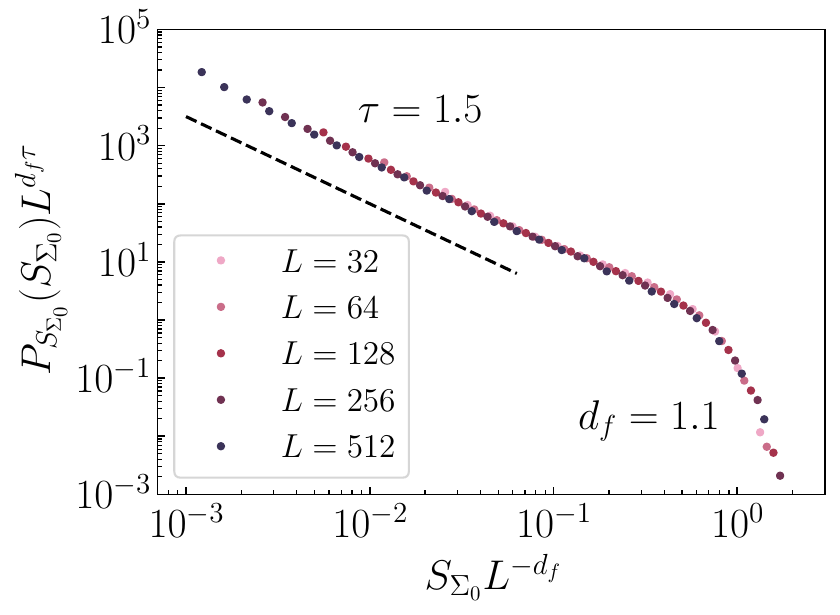}
    \caption{Finite-size scaling of the distribution of first avalanche sizes obtained by AQS starting from elastic states sampled by the random triggering protocol for several system sizes $L$.}
    \label{fig:finite-size-scaling-first-drop}
\end{figure}

We also perform the same analysis for the distribution of concatenated AQS avalanches, as discussed in Sec.~\ref{sec:control_cutoff}. 
Remarkably, we do find a collapse, as shown in \autoref{fig:finite-size-scaling-concatenated}, with $d_f \approx 1.5$ and $\tau \approx 1.5$.
Therefore concatenated avalanches still follow a critical scaling, albeit with different exponents from single avalanches in the AQS protocol.

In \autoref{fig:finite-size-scaling-first-drop}, we confirm the importance of the triggering mechanism via the way it controls the statistics of visited elastic states, by performing the finite-size scaling analysis on the distribution of first avalanches when we apply the AQS dynamics on elastic states sampled with the random triggering protocol.
Here again, a remarkable collapse is obtained. 
Looking at the exponents, we find $d_f \approx 1.1$ (like usual AQS avalanches) and $\tau = 1.5$ (unlike usual AQS avalanches). 
This shows that the elastic states sampled by the random triggering protocol are statistically different from the ones sampled by AQS, leading to different $\tau$ exponents.

\begin{figure}[t]
    \centering
    \includegraphics[width=0.8\linewidth]{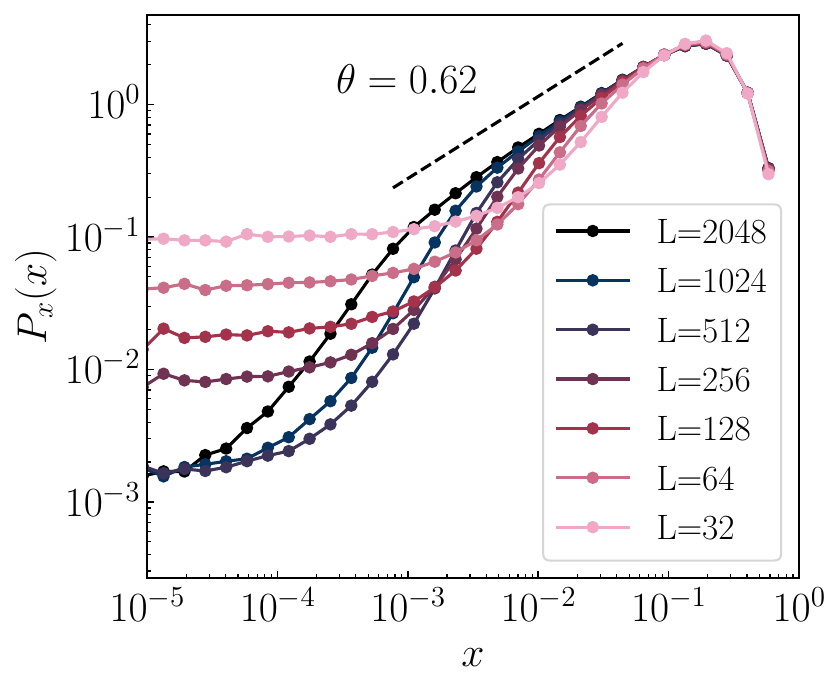}
    \caption{Distributions of the distance to threshold in the elastic states of the weakest triggering protocol for system sizes $L\in [32, 64, 128, 256, 512, 1024, 2048]$}
    \label{fig:p_x_weakest}
\end{figure}

\section{Pseudo-gap distribution in the weakest triggering protocol}
\label{sec:p_x_weakest}

In this appendix we report the size dependence of the distribution $P_x(x)$ of distance $x$ to local yield, for elastic states sampled using the weakest triggering protocol.
As shown in \autoref{fig:p_x_weakest}, this distribution only shows a  pseudo-gap $P_x(x)\sim x^\theta$ for the largest sizes $L$ that we are able to investigate numerically. 
This stands in contrast to the AQS and random triggering protocols, for which the pseudo-gap is clearly visible even for reasonable sizes.
Nonetheless, it appears in \autoref{fig:p_x_weakest} that the intermediate range of $x$ for which the scaling holds is widening, even if slowly,
when increasing system size. This suggests that in the large system size limit the pseudo-gap would actually be observed, with $\theta \approx 0.62$, the same value as for the AQS and random triggering protocols.

\end{document}